\documentclass[amsmath,amssymb,aps,eqsecnum]{revtex4}

\begin{document}
\title{Hilbert--Schmidt volume of the set of mixed quantum states}
\author{Karol
{\.Z}yczkowski$^{1,2}$ and Hans-J{\"u}rgen Sommers$^3$}
\affiliation {$^1$Instytut Fizyki im. Smoluchowskiego,
Uniwersytet Jagiello{\'n}ski,
ul. Reymonta 4, 30-059 Krak{\'o}w, Poland}
\affiliation{$^2$Centrum Fizyki Teoretycznej,
Polska Akademia Nauk, Al. Lotnik{\'o}w 32/44, 02-668 Warszawa, Poland}
\affiliation{$^3$Fachbereich Physik,
Universit\"{a}t Duisburg-Essen, Standort Essen, 
  45117 Essen, Germany}

 \date{July 9, 2003}

\begin{abstract}
We compute the volume of the convex $N^2-1$ dimensional set
 ${\cal M}_{N}$ of density matrices of size
$N$ with respect to the Hilbert-Schmidt measure.
The hyper--area of the boundary of this set is also found
and its ratio to the volume
provides an information about the complex structure of ${\cal M}_{N}$.
Similar investigations are also performed for the smaller set of all
real, symmetric density matrices. As an intermediate step
we analyze volumes of the unitary and orthogonal groups and of
the flag manifolds.

\end{abstract}


\maketitle

\medskip
\begin{center}
{\small e-mail: karol@cft.edu.pl \ \quad \
sommers@theo-phys.uni-essen.de}
\end{center}


\section{Introduction}

Although the notion of a density matrix
is one of the fundamental concepts
discussed in the elementary courses of quantum mechanics,
the structure of the set ${\cal M}_{N}$ of all
density matrices of size $N$ is not easy to characterize
\cite{Bl76,Ha78,ACH93}.
The only exception is the case $N=2$, for which
${\cal M}_{2}$ embedded in ${\mathbb R}^3$
has an appealing form of the {\sl Bloch ball}.
Its boundary, $\partial {\cal M}_{2}$,
consists of pure states and  forms the {\sl Bloch sphere}.
For larger number of states the dimensionality
of ${\cal M}_{N}$ grows quadratically with $N$, which makes
its analysis involved. In particular,
for $N>2$ the set of pure states forms
a $2N-2$ dimensional manifold, of measure zero
in the $N^2-2$ dimensional boundary $\partial {\cal M}_{N}$.

In this work we compute the volume of
${\cal M}_{N}$ with respect to the Hilbert--Schmidt (HS) measure.
The HS measure is defined by the HS metric which is distinguished by
the fact that it induces the flat, Euclidean geometry into
the set of mixed states. 
The (hyper)area of the boundary of the 
space of the density matrices, $\partial {\cal M}_{N}$,
is also computed, as well as the area of (hyper)edges of this set -
the HS volume of the
subspace of density matrices of an arbitrary rank $k <N$.
In the special case of $k=1$ we obtain a well--known formula
for the volume of the space of pure states,
equivalent to the complex projective manifold ${\mathbb C}P^{N-1}$.

A similar analysis is also performed
for the set of real density matrices. To calculate the
volume of the set of complex (real) mixed states we use
the volume of the unitary (orthogonal) groups and the volume of the
complex (real) flag manifolds - these results are described in the appendix.

A motivation for such a study is twofold.
On one hand, the complex structure of the set of mixed 
quantum states is interesting for itself. 
It is well--known that
for $N>2$ the $D=N^2-1$ dimensional set ${\cal M}_{N}$
is neither a $D$--ball nor a polytope, 
but, how it looks like? 
More like a ball or more like a polytope?
Instead of using techniques of differential
geometry and computing the average curvature
on the boundary of the set ${\cal M}_{N}$,
we compute the volume of its boundary 
and compare it with the volume  of the $D-1$ sphere,
which surrounds the ball of the same volume as ${\cal M}_{N}$.
Such a comparison shows us, to what extend the 
shape of the body of mixed quantum states differs from the ball,
in a sense that more (hyper)area of the surface
is needed to cover the same volume.

A complementary information characterizing 
the  structure of a given set 
is obtained by calculating the ratio between
the area of its boundary and its volume.
Among all $D$-dimensional bodies of a fixed volume,
such a ratio is smallest for the $D$--ball.
Hence computing such a ratio for the
$D$--dimensional body of mixed quantum states we may 
compare it with similar ratios obtained
for $D$--balls, $D$--cubes and $D$--simplices.

On the other hand our investigations might be useful
in characterizing the absolute volume of the subset
of mixed states distinguished by a certain attribute.
For instance, if $\varrho$ describes a composite system,
one may ask, what is the volume of the set of
separable (entangled) mixed states \cite{ZHSL98,Zy99}.
Furthermore, assume we are given a concrete mixed quantum state 
$\varrho$.
It is natural to ask, wether $\varrho$ is in some sense typical,
e.g. wether its von Neumann entropy is close to the average 
taken over the entire set ${\cal M}_{N}$
with respect to the HS measure. 
To compute such averages (see e.g. \cite{ZS01})
it is usefull to know the volume of ${\cal M}_{N}$
and to make use of integrals developed for such a 
calculation.

\section{Geometry of ${\cal M}_{N}$ with respect to the
Hilbert-Schmidt metric}

The set of mixed quantum states ${\cal M}_{N}$
consists of Hermitian, positive matrices of size $N$,
 normalized by the trace condition
\begin{equation}
{\cal M}_{N}:=\{\varrho:
\varrho=\varrho^{\dagger}; \ \
\varrho\ge 0; \ \
{\rm tr }\varrho=1;
\ \ {\rm dim}(\rho)=N \}.
  \label{setM}
\end{equation}
It is a compact convex set of dimensionality $D=N^2-1$.
Any density matrix may be diagonalized by a unitary
rotation,
\begin{equation}
\varrho=U \Lambda U^{-1},
\label{diagC}
\end{equation}
where $\Lambda$ is a diagonal matrix of eigenvalues $\Lambda_i$.
Due to the trace condition they satisfy
$\sum_{i=1}^{N}\Lambda_{i}=1$,
so the space of spectra is isomorphic with a $(N-1)$--dimensional
simplex $\Delta_{N-1}$.

Let $B$ be a diagonal unitary matrix. Since
$\varrho =UB\Lambda B^{\dagger }U^{\dagger}$,
in the generic case of a non degenerate spectrum
the unitary matrix $U$ is determined
up to $N$ arbitrary phases entering $B$.
To specify uniquely the unitary matrix of eigenvectors $U$
it is thus sufficient to select a point on the
coset space $Fl^{(N)}_{\mathbb C}:=U(N)/[U(1)]^N$, called complex {\sl
flag manifold}. The generic density matrix is thus determined by
$(N-1)$ parameters determining eigenvalues and
$N^2-N$ parameters related to eigenvectors,
which sum up to the dimensionality $D$ of ${\cal M}_{N}$.
Although for degenerated spectra the dimension of the flag
manifold decreases (see e.g. \cite{ACH93,ZSlo01}), these cases of measure zero do not influence the
estimation of the volume of the entire set of density
matrices.
Several different distances may be introduced
into the set  ${\cal M}_{N}$,
(see for instance \cite{PS96,ZSlo01}).
In this work we shall use the Hilbert-Schmidt metric,
which induces the flat geometry.

The Hilbert-Schmidt distance between any two density operators is
defined as the Hilbert-Schmidt (Frobenius) norm of their difference,
\begin{equation}
D_{\rm HS}(\varrho_1,\varrho_2)= ||\varrho_1 -\varrho_2||_{\rm HS} =
\sqrt{ {\rm Tr}   [(\varrho_1 - \varrho_2)^2] }.
  \label{HS1}
\end{equation}
The set of all mixed states of size two
acquires under this metric
the geometry of the Bloch ball ${\bf B}^3$ embedded in
${\mathbb R}^3$. Its boundary,
$\partial {\bf B}^3={\bf S}^2$ contains all pure states and is called
{\sl Bloch sphere}. To show this let us use the Bloch
representation of a $N=2$ density matrix
\begin{equation}
\varrho =  \frac {\mathbb I}{N} +
{\vec \tau}\cdot {\vec \lambda} \ ,
\label{Pauli}
\end{equation}
where $\vec{\lambda}$ denotes the vector of
three rescaled traceless Pauli matrices
$\{\sigma_x, \sigma_y, \sigma_z \}/{\sqrt{2}}$.
They are normalized according to tr$\lambda_i^2=1$.
The three dimensional Bloch vector $\vec \tau$ is real due to Hermiticity of
$\varrho$. Positivity requires  tr$\varrho^2\le 1$ and this
implies $|\vec \tau|\le 1/\sqrt{2}=:R_2$.
Demanding equality one distinguishes the
set of all pure states, $\varrho^2=\varrho$,
 which form the Bloch sphere of radius $R_2$.
Consider two arbitrary density matrices
and express their difference $\varrho_1-\varrho_2$
in the representation (\ref{Pauli}).
The entries of this difference consists of the differences
between components of both Bloch vectors ${\vec \tau}_1$  and
${\vec \tau}_2$. Therefore
\begin{equation}
  D_{\rm HS}\bigl( \varrho_{{\vec \tau}_1}, \varrho_{{\vec \tau}_2}\bigr)=
  D_{E}({\vec \tau}_1, {\vec \tau}_2)  \ ,
\label{densHS}
\end{equation}
where $D_E$ is the Euclidean distance between both
Bloch vectors in ${\mathbb R}^3$.
This proves that with respect to the HS metric the set ${\cal M}_{2}$
possesses the geometry of a ball ${\bf B}^3$ .
The unitary rotations of a density matrix
$\varrho \to U\varrho U^{\dagger}$
correspond to the rotations of $\vec \tau$ in ${\mathbb R}^3$.
This is due to the fact that the adjoint representation of
$SU(2)$ is isomorphic with $SO(3)$.

The Hilbert--Schmidt metric induces a flat geometry inside
${\cal M}_{N}$ for arbitrary $N$.
  Any state $\varrho$ may be represented by
(\ref{Pauli}), but now the $\vec \lambda$
represents an operator valued vector which
consists of $D=N^2-1$ traceless Hermitian generators of $SU(N)$,
which fulfill tr$\lambda_i\lambda_j=\delta_{ij}$.
This generalized Bloch representation
of density matrices for arbitrary $N$ was introduced
by Hioe and Eberly \cite{HE81},
and recently used in \cite{RMNDMC01}.
The case $N=3$, related to the Gell-Mann matrices,
is discussed in detail in the paper by Arvind et al. \cite{AMM97}.
The generalized Bloch vector  $\vec \tau$
(also called {\sl coherence vector}) is
$D$ dimensional. In the general case
of an arbitrary $N$ the right hand side of (\ref{densHS})
denotes the Euclidean distance between two Bloch vectors
in ${\mathbb R}^{N^2-1}$.
Positivity of $\rho$ implies the bound for its length
\begin{equation}
 |\vec{\tau}| \le D_{\rm HS}\bigl(
{\mathbb I}/N ,|\psi\rangle \langle {\psi}| \bigr)
 = \sqrt{\frac{N-1}{N}}=:R_N .
\label{Rn}
\end{equation}

In contrast to the Bloch sphere,
the complex projective space
${\mathbb C}{\bf P}^{N-1}$ which contains all pure states,
forms for $N>2$ only a measure zero, simply connected
$2(N-1)$-dimensional subset of the
$N^2-2$ dimensional sphere  of radius $R_N$
embedded in  ${\mathbb R}^{N^2-1}$.
Thus not every vector $\vec \tau$ of the maximal length  $R_N$
represents a quantum state.
This is related with the fact that for $N\ge 3$ the adjoint
representation of $SU(N)$ forms only a subset of $SO(N^2-1)$,
 (see e.g. \cite{Ma68}).
Sufficient and necessary conditions
for a Bloch vector to represent a pure state were
given in  \cite{AMM97} for $N=3$, and
in \cite{JS01} for an arbitrary $N$.
Furthermore, by far not all vectors of length shorter then
$R_N$ represent a quantum state,
as not all the points inside a hyper-sphere belong to the simplex
inscribed inside it. Necessary conditions for a 
Bloch vector to represent quantum mixed state 
were recently provided by Kimura \cite{Ki03}.
On the other hand, there exists a smaller sphere
inscribed inside the  set ${\cal M}_{N}$.
Its radius  reads \cite{Ha78}
\begin{equation}
 r_N = D_{\rm HS}\bigl({\mathbb I}/N ,\varrho_{N-1} \bigr)
\frac{ 1}{\sqrt{N(N-1)}}=\frac{R_N}{N-1},
\label{radius2}
\end{equation}
where $\varrho_{N-1}$ denotes any state with the spectrum
$(\frac{1}{N-1},...,\frac{1}{N-1},0)$.

\section{Hilbert-Schmidt measure}

Any  metric in the space of mixed quantum states generates
a measure, inasmuch as one can assume that
drawing random density matrices from each ball of a fixed radius is
equally likely. The balls are understood with respect to a given metric.
In this work we investigate the measure induced by the Hilbert-Schmidt
distance (\ref{HS1}).
The infinitesimal distance takes a particularly simple form
\begin{equation}
 ({\rm d}s_{\rm HS})^2=  {\rm Tr}  [({\rm d} \varrho)^2]
  \label{HS2}
\end{equation}
valid for any dimension $N$.
Making use of the diagonal form $\rho=U\Lambda U^{-1}$
we may write
\begin{equation}
 {\rm d}\varrho = U[ {\rm d}\Lambda +U^{-1}{\rm d U}\Lambda -
  \Lambda U^{-1}{\rm d U} ] U^{-1}.
  \label{drho1}
\end{equation}
Thus (\ref{HS2}) can be rewritten as
\begin{equation}
 ({\rm d}s_{\rm HS})^2=  \sum_{i=1}^N ({\rm d}\Lambda_i)^2 +
   2 \sum_{i<j}^N (\Lambda_i-\Lambda_j)^2 |(U^{-1}{\rm d}U)_{ij}|^2 .
  \label{HS2b}
\end{equation}
Since the density matrices are normalized,
$\sum_{i=1}^N \Lambda_i=1$, thus
$\sum_{i=1}^N {\rm d}\Lambda_i=0$.
Hence one may consider the variation of the $N$-th eigenvalue
as a dependent one, ${\rm d}\Lambda_N=
-\sum_{i=1}^{N-1} {\rm d}\Lambda_i$,
which implies
\begin{equation}
 \sum_{i=1}^N ({\rm d}\Lambda_i)^2 =
 \sum_{i=1}^{N-1} ({\rm d}\Lambda_i)^2 +
 \bigl( \sum_{i=1}^{N-1} {\rm d}\Lambda_i \bigr)^2 =\sum_{i,j=1}^{N-1}{\rm d}\Lambda_i g_{ij}{\rm d}\Lambda_j .
  \label{HS2c}
\end{equation}
The corresponding volume element
gains a factor $\sqrt{{\rm det}g}$,
where $g$ is the metric in
 the $(N-1)$ dimensional simplex $\Delta_{N-1}$
of eigenvalues. From (\ref{HS2c})
one may read out the explicit form of the metric
$g_{ij}$
\begin{equation}
g=\left[ \begin{array}  [c]{ccc}
1 &  & 0\\
& \ddots & \\
0 &  & 1
\end{array} \right]  +
\left[ \begin{array}
[c]{ccc}
1 & \cdots & 1\\
\vdots & \ddots & \vdots\\
1 & \cdots & 1
\end{array} \right]   \ .
  \label{metricg}
\end{equation}
It is easy to check that the spectrum of the $N-1$ dimensional matrix
$g$ consists of one eigenvalue equal to $N$ and remaining $N-2$
eigenvalues equal to unity, so that det$g=N$.
Thus the Hilbert-Schmidt volume element is given by
\begin{equation}
{\rm d}V_{\rm HS} = \sqrt{N} \prod_{j=1}^{N-1} {\rm d}\Lambda_j
     \prod_{j<k}^{1\cdots N}
   (\Lambda_j-\Lambda_k)^2\
     | \prod_{j<k}^{1\cdots N}
    2 {\rm Re}(U^{-1}{\rm d}U)_{jk}
       {\rm Im}(U^{-1}{\rm d}U)_{jk} | .
  \label{HS2d}
\end{equation}
and has the following product form
\begin{equation}
{\rm d} V = {\rm d} \mu (\Lambda_1,\Lambda_2,...,\Lambda_N) \times 
{\rm d} \nu_{\rm Haar} \ .
  \label{dVdmupr}
\end{equation}
The first factor
depends only on the eigenvalues $\Lambda_i$,  while the
latter on the eigenvectors of $\varrho$ which compose the unitary matrix $U$.

Any unitary matrix may be considered as an element of the
Hilbert-Schmidt space of operators
with the scalar product
$\langle A|B\rangle={\rm Tr}A^{\dagger}B$.
This suggests the following definition of an invariant metric of the unitary group $U(N)$,
\begin{equation}
({\rm d}s)^2  := -{\rm Tr} (U^{-1} {\rm d}U)^2=
\sum_{jk=1}^N|(U^{-1}{\rm d}U)_{jk}|^2 =
\sum_{j=1}^N|(U^{-1}{\rm d}U)_{jj}|^2 +
2 \sum_{j<k=1}^N |(U^{-1}{\rm d}U)_{jk}|^2 \ .
\label{ds2}
\end{equation}
This metric induces the unique Haar measure $\nu_{Haar}$ on $U(N)$,
invariant with respect to unitary
transformations, $\nu_{\rm Haar}(W)=\nu_{\rm Haar}(UW)$, where $W$ denotes an arbitrary
measurable subset of $U(N)$.
Integrating the volume element corresponding to (\ref{ds2}) over the unitary group we obtain the
volume
\begin{equation}
 {\rm Vol} \bigl[ U(N)\bigr] =
\frac{ (2\pi)^{N(N+1)/2}}{
 1! 2! \cdots (N-1)!} .
\label{volUNb}
\end{equation}
Integrating the volume element  with the diagonal terms in (\ref{ds2}) omitted (in that case the diagonal elements of $U$ are fixed by $U_{ii}\ge 0$)
we obtain the volume of the complex flag manifold,
$Fl^{(N)}_{\mathbb C}:=U(N)/[U(1)^N]$,
\begin{equation}
 {\rm Vol} \bigl[ Fl^{(N)}_{\mathbb C}\bigr] =
\frac{ {\rm Vol} \bigl[ U(N)\bigr]}{ (2\pi)^N} =
\frac{ (2\pi)^{N(N-1)/2}}{ 1! 2! \cdots (N-1)!} .
\label{volFlb}
\end{equation}
Both results are known in the literature for almost fifty years
\cite{Hu63}. However, since many different conventions in defining the
volume of the unitary group are in use
\cite{Ma81,Tu85,Fu01,TBS02,BST02,TS02,Ca02}
we sketch a derivation of the above expressions
in the appendix and provide a list of related results.

Comparing formulae (\ref{HS2d}) and (\ref{ds2})
we recognize that
the measure $\nu$, responsible for the choice of
eigenvectors of $\varrho$, is the natural measure on the complex flag
manifold $Fl^{(N)}_{\mathbb C}=U(N)/[U(1)^N]$ induced by the Haar
measure on $U(N)$.
Since the trace is unitarily invariant, it follows directly from the
definition (\ref{HS2}),
that the volume element with respect to the HS measure
is invariant with respect to the group of
unitary rotations, ${\rm d}V_{\rm HS}(\varrho)={\rm d}V_{\rm HS}(U\varrho U^{\dagger})$.
Such a property is characteristic to any
{\sl product measure} of the form (\ref{dVdmupr}).
Several product measures with different choices of $\mu$
were examined in \cite{Zy99,Sl99a,ZS01,Ca02}.

Integrating the volume element (\ref{HS2d})
with respect to the eigenvectors of $\varrho$ distributed according to
 the Haar measure one obtains the
probability distribution in the simplex of eigenvalues
\begin{equation}
 P^{(2)}_{\rm HS}(\Lambda_1,\dots,\Lambda_N) = C^{\rm HS}_N
\delta(1-\sum_{j=1}^N
\Lambda_j)  \prod_{j<k}^N (\Lambda_j-\Lambda_k)^2,
  \label{HS3}
\end{equation}
where for future convenience we have decorated the symbol $P$ with the
superscript $^{(2)}$ consistent with the exponent in the last factor.
 As discussed in the following
section the normalization constant $C^{\rm HS}_N$
may be expressed \cite {ZS01} in terms of the Euler
Gamma function $\Gamma(x)$ \cite{SO87}
\begin{equation}
 C_{N}^{\rm HS} =  \frac{\Gamma(N^2)}
{\prod_{j=0}^{N-1} \Gamma(N-j) \Gamma(N-j+1) } .
\label{constn}
\end{equation}
The above joint probability distribution, derived by Hall \cite{Ha98},
defines the measure $\mu_{\rm HS}$ in the space of
diagonal matrices, and the
 {\sl Hilbert-Schmidt} measure (\ref{HS2d}) in the space of
 density matrices ${\cal M}_{N}$.

Interestingly, the very same measure may be generated by
drawing random pure states $|\phi\rangle\in {\cal H}_1 \otimes {\cal
H}_2$ of a composite $N \times N$ system
according to the Fubini-Study measure on ${\mathbb C}{\bf P}^{N^2-1}$.
Then the density matrices of size $N$ obtained by partial trace,
$\varrho={\rm tr}_2(|\phi\rangle\langle\phi|)$,
are distributed according to the HS measure \cite{Br96,Ha98,ZS01}.
Alternatively, one may generate a random matrix $A$ of the Ginibre
ensemble, (non-Hermitian complex matrix with all entries independent
Gaussian variables with zero mean and a fixed variance)
and obtain a HS distributed random density matrix
by a projection $\varrho=A^{\dagger}A/{\rm tr}A^{\dagger}A$
 \cite{ZS01}. A similar approach was recently advocated by Tucci
\cite{Tu02},
who used the name 'uniform ensemble' just for ensemble of density
matrices generated according to the HS measure.

\section{Volume of the set of mixed states}

For later convenience let us introduce generalized normalization constants
\begin{equation}
\frac{1}{C_{N}^{(\alpha,\beta)}}:=
\int_0^{\infty} {\rm d}\Lambda_1 \cdots {\rm d}\Lambda_N
\delta(\sum_{i=1}^N \Lambda_i -1) \prod_{i=1}^N \Lambda_i^{\alpha-1}
\prod_{i<j} |\Lambda_i-\Lambda_j|^{\beta}
\label{constab}
\end{equation}
with $\alpha, \beta >0$.
These constants may be calculated using the formula for the Laguerre
ensemble, discussed in the book of Mehta \cite{Me91},
\begin{equation}
\int_0^{\infty} {\rm d}\Lambda_1 \cdots {\rm d}\Lambda_N
\exp\bigl(-\sum_{i=1}^N \Lambda_i\bigr)
 \prod_{i=1}^N \Lambda_i^{\alpha-1}
\prod_{i<j} |\Lambda_i-\Lambda_j|^{\beta}
= \prod_{j=1}^N \Bigl[
\frac{ \Gamma[1+j\beta/2] \Gamma[\alpha +(j-1)\beta/2]}
     { \Gamma[1+\beta/2]} \Bigr] .
\label{constab1}
\end{equation}
Substituting $x_i^2=\Lambda_i$ we may bring the latter
integral to the Gaussian form. Expressing it in the spherical
coordinates we get the integral (\ref{constab})
and eventually obtain
\begin{equation}
\frac{1}{C_N^{(\alpha,\beta)} }:=
\frac{1}{\Gamma[\alpha N +\beta N(N-1)/2]}
\prod_{j=1}^N \Bigl[
\frac{ \Gamma[1+j\beta/2] \Gamma[\alpha +(j-1)\beta/2]}
     { \Gamma[1+\beta/2]} \Bigr] .
\label{constab2}
\end{equation}
By definition $C_N^{\rm HS}=C_N^{(1,2)}$ and the special case of
the above expression reduces to (\ref{constn}).

To obtain the Hilbert-Schmidt volume of the
set of mixed states ${\cal M}_{N}$
one has to integrate the volume element (\ref{HS2d})
over eigenvalues and eigenvectors.
By definition the first integral gives $1/C_N^{\rm HS}$,
while the second is equal to the volume of the flag manifold.
To make the diagonalization transformation (\ref{diagC})
unique one has to restrict to a certain order of eigenvalues,
say,
$\Lambda_1 < \Lambda_2 < \cdots < \Lambda_N$,
(a generic density matrix is not degenerate),
which corresponds to a choice of
a certain Weyl chamber of the eigenvalue simplex  $\Delta_{N-1}$.
In other words,  different permutations
 of the vector of $N$ generically different
eigenvalues $\Lambda_i$ belong to the same unitary orbit.
The number of different permutations (Weyl chambers) equals to
$N!$, so the volume reads
\begin{equation}
V^{(2)}_N:={\rm Vol}_{\rm HS} \bigl( {\cal M}_{N} \bigr) =
\frac{\sqrt{N}}{N!} \
\frac{{\rm Vol}\bigl(Fl^{(N)}_{\mathbb C}\bigr)}{C_N^{\rm HS} }.
\label{volmix2a}
\end{equation}
The square root stems from the volume element (\ref{HS2d}),
and the index $^{(2)}$ refers to the general case of complex
density matrices. Making use of (\ref{constn}) and (\ref{volFlb})
we arrive at the final result
\footnote {Apart of the first factor $\sqrt{N}$,
the same formula appeared already in  the work of Tucci
\cite{Tu02}}
\begin{equation}
V^{(2)}_N
=\sqrt{N} (2\pi)^{N(N-1)/2}\ \frac{\Gamma(1) \cdots \Gamma(N)}
{\Gamma(N^2)} .
\label{volmix2b}
\end{equation}

Substituting $N=2$ we are pleased to receive
$V^{(2)}_2=\pi \sqrt{2}/3$ - exactly the volume of the Bloch ball
${\bf B}^3$ of radius $R_2=1/\sqrt{2}$.
This result may be also found in the
notes by Caves \cite{Ca02}, who also derived 
an explixit integral for the volume 
of the set of mixed states for arbitrary $N$.

The next result $V^{(2)}_3=\pi^3 /(840\sqrt{3})$
allows us to characterize the difference
between the set ${\cal M}_{3}\subset {\mathbb R}^8$ and the ball ${\bf
B}^8$. The set of mixed states is inscribed into the sphere of radius
$R_3=\sqrt{2/3}\approx 0.816$, while
the maximal ball contained inside has the radius
$r_3=R_3/2\approx 0.408$. Using  Eq.
(\ref{volSNBN}) we find the radius of the $8$--ball
of volume $V_3$ is $\rho_3\approx 0.519$.
The distance from the center of ${\cal M}_{3}$ to its boundary
varies with the direction in ${\mathbb R}^8$
from $r_3$ to $R_3$,
in contrast to the $N=2$ case of Bloch ball,
for which $R_2=r_2=\rho_2=1/\sqrt{2}$.
The average HS--distance from the center of
${\cal M}_{3}$ to its boundary  is equal to
$\rho_3$.
Similar calculations performed
for $N=4$ give the maximal radius $R_4=\sqrt{3/4}\approx 0.866$,
the minimal radius $r_4=R_4/3\approx 0.289$ and the 'mean'
radius $\rho_4\approx 0.428$ which
generates the ball ${\bf B}^{15}$ of the same volume as $V_4$.
In general, let $\rho_N$ denote the radius of a
ball  ${\bf B}^{N^2-1}$ of the same volume as
the set ${\cal M}_{N}$.

The volume $V_N$ tends to zero if $N\to \infty$, but there
is no reason to worry about it. The same is true for the
volume of the $N$--ball, see (\ref{volSNBN}).
This is just a consequence of the choice of the units.
We are comparing the volume of an object in ${\mathbb R}^N$
with the volume of a hypercube $C^N$ of side one,
and it is easy to understand, that the larger dimension, the smaller is
the volume of the ball inscribed into it.

\section{Area of the boundary of the set of mixed states}

The boundary of the set of mixed states is far from being trivial.
Formally it may be written as a solution of the equation
det$\varrho=0$ which contains all matrices of a lower rank.
The boundary $\partial {\cal M}_{N}$
 contains orbits of different dimensionality
generated by spectra of different rank and degeneracy
(see eg. \cite{ACH93,ZSlo01}). Fortunately all
of them are of measure zero besides the
generic orbits created by unitary rotations of diagonal matrices with
all eigenvalues different and one of them equal to zero;
$\Lambda=\{0, \Lambda_2 <\Lambda_3 <\cdots <\Lambda_N\}$.
Such spectra form the $N-2$ dimensional
simplex $\Delta_{N-2}$, which contains $(N-1)!$ Weyl chambers -
this is the number of possible permutations
of elements of $\Lambda$ which all belong to the same unitary orbit.

Hence the hyper-area of the boundary may be computed
in a way analogous to (\ref{volmix2a}),
\begin{equation}
S^{(2)}_{N}:={\rm Vol}_{\rm HS} \bigl( {\cal M}_{N} \bigr) =
\frac{\sqrt{N-1}}{(N-1)!} \
\frac{{\rm Vol}\bigl(Fl^{(N)}_{\mathbb C}\bigr)}{C_{N-1}^{(3,2)}
}.
\label{volare2a}
\end{equation}
The change of the parameter $\alpha$ in (\ref{constab}) from $1$ to $3$
is due to the fact that by setting one component of an $N$ dimensional
  vector to zero the corresponding Vandermonde determinant
of size $N$ leads to   the determinant of size $N-1$ for $\beta = 1$
and to the square of the determinant for $\beta = 2$.
Applying (\ref{constn}) and (\ref{volFlb})
we obtain an explicit result
\begin{equation}
S^{(2)}_{N}=
\sqrt{N-1} \ (2\pi)^{N(N-1)/2}\ \frac{\Gamma(1) \cdots \Gamma(N+1)}
{\Gamma(N)\Gamma(N^2-1)} .
\label{volare2b}
\end{equation}
For $N=2$ we get $S_2^{(2)}=2\pi$ - just the area of the
Bloch sphere ${\bf S}^2$ of radius  $R_2=1/\sqrt{2}$.
The area of the $7$-dim  boundary of ${\cal M}_{3}$
reads $S_3^{(2)}=\sqrt{2} \pi^3/105$.

In an analogous way we may find the volume of edges,
formed by the unitary orbits of the vector of eigenvalues with two
zeros. More generally, states of rank $N-n$ are unitarily similar to
diagonal matrices with $n$ eigenvalues vanishing,
$\Lambda=\{0,\dots,0,\Lambda_{n+1} <\Lambda_{n+2} <\cdots
<\Lambda_N\}$. These edges of order $n$ are $N^2-n^2-1$
dimensional, since the dimension of the
set of such spectra is $N-n-1$, while
the orbits have the structure of $U(N)/[U(n)\times (U(1))^{N-n}]$
and dimensionality $N^2-n^2-(N-n)$.
Repeating the reasoning used to derive (\ref{volare2a})
we obtain the volume of the hyperedges
\begin{equation}
S^{(2)}_{N,n}=
\frac{\sqrt{N-n}}{(N-n)!} \
\frac{1}{ C_{N-n}^{(1+2n,2)}} \
\frac{{\rm Vol}\bigl(Fl^{(N)}_{\mathbb C}\bigr)}
     { {\rm Vol}\bigl(Fl^{(n)}_{\mathbb C}\bigr)}.
\label{voledges}
\end{equation}
Note that for $n=0$ this expression gives the volume
$V_N^{(2)}$ of the set ${\cal M}_{N}$, for $n=1$ the hyperarea
$S_N^{(2)}$ of its boundary $\partial {\cal M}_{N}$,
for $n\ge 2$ the area of the edges of rank $N-n$. In the extreme case
of $n=N-1$ the above formula gives correctly the volume of the set of
pure states, (the states of rank one),
Vol$({\mathbb C}{\bf P}^{N-1})=(2\pi)^{N-1}/\Gamma(N)$, see appendix.

\section{The ratio:  area/volume}

Certain information about the structure of a convex body
may be extracted from the ratio $\gamma$
of the (hyper)area of its boundary to its volume.
The smaller the coefficient $\gamma$ (with the diameter of the body kept 
fixed), the better the body investigated may be approximated by a ball,
for which such a ratio is minimal. And conversely, the larger 
$\gamma$, the less the body resembles a ball,
since more (hyper)area is needed to bound a given volume.

To analyze simple examples let us recall the
 volume of the $N$-dimensional unit ball ${\bf B}^N\subset {\mathbb
R}^{N}$
 and the volume $S_N$ of the unit $N$--sphere
${\bf S}^N\subset {\mathbb R}^{N+1}$
\begin{equation}
 B_N:=\mbox{vol}({\bf B}^N) = \frac{\mbox{vol}({\bf S}^{N-1})}{N} =
\frac{{\pi}^{\frac{N}{2}}}{{\Gamma}(\frac{N}{2} + 1)}
\sim
\frac{1}{\sqrt{2{\pi}}}
\left( \frac{2{\pi}e}{N}\right)^{\frac{N}{2}} ,
\label{volSNBN}
 \end{equation}
where  the Stirling expansion \cite{SO87} was used for large $N$.
For small $N$ we obtain the well known expressions,
$S_1=2\pi$, $S_2=4\pi$, $S_3=2\pi^2$, $S_4=8\pi^2/3$ and
$B_2=\pi$, $B_3=4\pi/3$, $B_4=\pi^2/2$.
If the spheres and balls have radius $L$ then the scale factor $L^N$
has to be supplied.
In odd dimensions the volume of the sphere simplifies,
${\mbox{vol}} ({\bf S}^{2k-1}) =2\pi^k/(k-1)!$.

Since the boundary of a $N$--ball
is formed by a $N-1$ sphere,
$\partial {\bf B}^N= {\bf S}^{N-1}$,
the ratio $\gamma$ for a ball of radius $L$ reads
\begin{equation}
\gamma( {\bf B}^N) :=
 \frac{\mbox{vol}({\bf \partial B}^N)} {\mbox{vol}({\bf B}^N)}
 = \frac{N}{L} .
\label{muball}
 \end{equation}
Intuitively this ratio will be
the smallest possible among all $N$-dimensional sets
of the same volume.
Hence let us compare it with an analogous result for a
hypercube
$\square_N$ of side $L$ and volume $L^N$. The cube has $2^N$
corners and $2N$ faces, of area $L^{N-1}$ each. We find
the ratio
\begin{equation}
\gamma( { \square}_N) :=
 \frac{\mbox{vol}({ \partial \square}_N)} {\mbox{vol}({\square}_N)}
 = 2 \ \frac{N}{L}
\label{mucube}
 \end{equation}
which grows twice as fast as for $N$-balls.
Another comparison can be made with simplices $\triangle_N$,
generated by $(N+1)$ equally distant points in ${\mathbb R}^N$.
The simplex $\triangle_2$ is a equilateral triangle, while
$\triangle_3$ is a regular tetrahedron.
The  volume of a simplex of side $L$  reads
$\mbox{vol}( \triangle_N) = [L^N \sqrt{(N+1)/2^N}]/N!$.
Since the boundary of $\triangle_N$ consists
of $N+1$ simplices $\triangle_{N-1}$ we obtain
\begin{equation}
\gamma( {\bf \triangle}_N) :=
 \frac{\mbox{vol}( \partial \triangle_N)}
{ {\mbox{vol}}(\triangle_N)}  =
\sqrt{\frac {2N}{N+1}}  \frac {N(N+1)}{L} \ .
\label{musimplex}
 \end{equation}
In this case the ratio $\gamma$
grows quadratically with $N$,
which reflects the fact that simplices do have
much 'sharper' corners, in contrast to the cubes, so more
(hyper)area of the boundary is required to cover a given volume.
Furthermore, if one defines a hyper--diamond as two simplices glued
along one face, its volume is twice the volume of $\triangle_N$ while
its boundary consists of $2N$ simplices $\triangle_{N-1}$, so the
coefficient $\gamma$ grows exactly as $N^2$.

Interestingly, the ratio $\gamma$ of the $N$--cube is the same
as for the $N$--ball inscribed in, which has much smaller volume.
The same property is characteristic for the $N$--simplex.
Hence another possibility to characterize the shape of any convex
body $F$ is to compute the ratio
 $\chi_1:={\rm vol}[{\bf B}_1(F)]/{\rm vol}(F)$,
and
 $\chi_2:={\rm vol}(F)/{\rm vol}[{\bf B}_2(F)]$,
where ${\bf B}_1(F)$ is the largest ball inscribed in $F$
while ${\bf B}_2(F)$ is the smallest ball in which
$F$ may be inscribed.
As stated above for cubes and simplices one has
$\gamma(F)=\gamma[{\bf B}_1(F)]$.

Such quotients may be computed for the rather complicated
convex body of mixed quantum states analyzed with respect to the
Hilbert-Schmidt measure.
Using expressions (\ref{volmix2a})
and (\ref{volare2a}) we find
\begin{equation}
\gamma_N:=
 \frac{\mbox{vol} \bigl( \partial {\cal M}_{N}  \bigr)}
{ {\mbox{vol}}    \bigl( {\cal M}_{N}  \bigr)}  =
\frac{N! \sqrt{N-1}}{\sqrt{N} (N-1)!} \
\frac{C_N^{(1,2)} } {C_{N-1}^{(3,2)}}=
\sqrt{N(N-1)}\ (N^2-1).
\label{muMN}
 \end{equation}
The first coefficients read
$\gamma_2=3 \sqrt{2}$, $\gamma_3=8\sqrt{6}$,
and $\gamma_4=15 \sqrt{12}$, so
they grow with $N$ faster than $N^{2}$.
A direct comparison with the results received for balls or cubes would
be unfair, since here $N$ does not denote the dimension of
the set ${\cal M}_{N}\subset {\mathbb R}^D$. Substituting the right
dimension, $D=N^2-1$, we see that
the area/volume ratio for the mixed states increases with
the dimensionality as $\gamma \sim D^{3/2}$.
The linear scaling factor $L$, equal to the radius $R_N$
tends asymptotically to unity and does not influence this behaviour.

Note that the set of mixed states is convex
and is inscribed into the sphere of radius $R_N$,
so for each finite $N$ the ratio $\gamma_N$
remains finite. On the other hand, the fact that this coefficient
 increases with the dimension $D$
much faster than for balls or cubes,
sheds some light into the intricate structure of the set
 ${\cal M}_{N}$.
It touches the hypersphere ${\bf S}^{N^2-2}$ of radius $R_N$
along the $2N-2$ dimensional manifold of pure states.
However, to be characterized by such a value of the
coefficient $\gamma$ it is a rather 'thin' set,
and a lot of hyper--area of the boundary is used to encompass its
volume. In fact, for any mixed state $\varrho \in {\cal M}_{N}$ its
distance to the boundary $\partial {\cal M}_{N}$
does not exceed the radius $r_N\sim 1/N$.
Another comparison can be made with the $D$-ball of radius
$L=r_N=[N(N-1)]^{-1/2}$, inscribed into
${\cal M}_{N}$. Although its volume is much smaller than this of the
larger set of mixed states, its area to volume ratio,
$\gamma=D/L$ is exactly equal to (\ref{muMN}) characterizing
${\cal M}_{N}$. In other words, for any dimensionality $N$ the set
of mixed quantum states belongs to the class of bodies
for which $\gamma(F)=\gamma({\bf B}_1(F))$ holds.

Using the notion of the effective radius $\rho_N$,
introduced in section IV,
we may express the coefficients $\chi_i$
for the set ${\cal M}_{N}$
as a ratio between radii raised to the power
equal to the dimensionality, $D=N^2-1$.
The exact values of
$\chi_1=(r_N/\rho_N)^D$ and $\chi_2=(\rho_N/R_N)^D$,
as well as their product $\chi=\chi_1\chi_2$,
may be readily obtained from (\ref{volmix2b}).
Let us only note the large $N$ behaviour,
$\chi( {\cal M}_{N}) =(N-1)^{-N^2+1}$
so it grows with the dimensionality $D$ as $D^{-D/2}$
while
$\chi({\bf B}^N)=1$,
 $\chi( {\square}_N)=N^{-N/2}$, and
 $\chi(\triangle_N)\approx N^{-N}$.

\section{Rebits: real density matrices}

Even though from physical point of view
one should in general consider the entire set ${\cal M}_{N}$
 of complex density matrices, we propose now to discuss its proper
subset: the set of real density matrices.
This set, denoted by ${\cal M}^{\mathbb R}_{N}$,
is of smaller dimension $D_1=N(N+1)/2-1<D=N^2-1$,
and any reduction of dimensionality simplifies the investigations.
While complex density matrices of size two are known as qubits,
the real density matrices are sometimes called {\sl rebits}
\cite{CFR02}. In the sense of the HS metric the space of rebits
forms the full circle  ${\bf B}^2$, which may be obtained as a slice of the
Bloch ball ${\bf B}^3$ along a plane containing ${\mathbb I}/2$.

To find the volume of the set ${\cal M}^{\mathbb R}_{N}$
we will repeat the steps (\ref{drho1})--(\ref{volmix2b})
for real symmetric density matrices which may be diagonalized
by an orthogonal rotation, $\varrho=O\Lambda O^T$.
The expressions
\begin{equation}
 {\rm d}\varrho = O[ {\rm d}\Lambda +O^{-1}{\rm d O}\Lambda -
  \Lambda O^{-1}{\rm d O} ] O^{-1}
  \label{drhoRe1}
\end{equation}
and
\begin{equation}
 ({\rm d}s_{\rm HS})^2=  \sum_{i=1}^N ({\rm d}\Lambda_i)^2 +
   2 \sum_{i<j}^N (\Lambda_i-\Lambda_j)^2 |(O^{-1}{\rm d}O)_{ij}|^2
  \label{HSRe2b}
\end{equation}
allow us to obtain the HS volume element, analogous to
(\ref{HS2d}),
\begin{equation}
{\rm d}V_{\rm HS}^{(1)} = \sqrt{N} \prod_{j=1}^{N-1} {\rm d}\Lambda_j
     \prod_{j<k}^{1\dots N}
    |\Lambda_j-\Lambda_k|\ \cdot
     \ |\prod_{j<k}^{1\cdots N}
     \sqrt{2}\bigl( O^{-1} {\rm d }O)_{jk}|
  \label{HS2Red}
\end{equation}
As in the complex case the measure has the product form, and
the last factor is the volume element of the orthogonal group (see
appendix).
Orthogonal orbits of a nondegenerate diagonal matrix
 form real flag manifolds
$Fl^{(N)}_{\mathbb R}=O(N)/[O(1)]^N$ of the volume
\begin{equation}
 {\rm Vol} \bigl[Fl^{(N)}_{\mathbb R}\bigr] =
\frac{ {\rm Vol} \bigl[ O(N)\bigr]}{ 2^N} =
\frac{ (2\pi)^{N(N-1)/4} \pi^{N/2}}{\Gamma[1/2]\cdots \Gamma[N/2]} \ .
\label{volFlReb}
\end{equation}
Here $O(1)$ is the reflection group ${\mathbb Z}_2$ with volume $2$.

The volume element (\ref{HS2Red})
leads to the following probability measure in the simplex of eigenvalues
\begin{equation}
 P^{(1)}_{\rm HS}(\Lambda_1,\dots,\Lambda_N) = C_N^{(1,1)}
\delta(1-\sum_{j=1}^N
\Lambda_j)  \prod_{j<k}^N |\Lambda_j-\Lambda_k|,
  \label{HSRe3}
\end{equation}
with the normalization constant given in (\ref{constab2}).
Note the linear dependence on the differences of eigenvalues, in
contrast to the quadratic form present in (\ref{HS3}).
Taking into account the number $N!$
of different permutations of the elements of the spectrum $\Lambda$
we obtain the expression for the volume of the set of
 ${\cal M}^{\mathbb R}_{N}$,
\begin{equation}
V^{(1)}_N:={\rm Vol}_{\rm HS} \bigl( {\cal M}^{\mathbb R}_{N} \bigr) =
\frac{\sqrt{N}}{N!} \
\frac{{\rm Vol}\bigl(Fl^{(N)}_{\mathbb R}\bigr)}{C_N^{(1,1)}} \ ,
\label{volmixRe2a}
\end{equation}
which gives
\begin{equation}
V^{(1)}_N
=\frac{\sqrt{N}}{N!}\
 \frac{2^N (2\pi)^{N(N-1)/4}\ \Gamma\bigl[ \frac{N+1}{2} \bigr] }
  {\Gamma\bigl[\frac{N(N+1)}{2}\bigr]\  \Gamma\bigl[\frac{1}{2}\bigr]}\
  \prod_{k=1}^N \Gamma\bigl[1+\frac{k}{2}]  \ .
\label{volmixRe2b}
\end{equation}

As in the complex case we find the
volume of the boundary of ${\cal M}^{\mathbb R}_{N}$,
and in general, the volume of edges of order $n$ with
$0\le n\le N-1$.  In the case of real density matrices
these edges are  $N(N+1)/2-1-n(n+1)/2$
dimensional, since the dimension of the
set of such spectra is $N-n-1$, and
the orbits have the structure of $O(N)/[O(n)\times (O(1))^{N-n}]$
and dimensionality $N(N-1)/2-n(n-1)/2$.
In analogy to (\ref{voledges}) we obtain
\begin{equation}
S^{(1)}_{N,n}=
\frac{\sqrt{N-n}}{(N-n)!} \
\frac{1}{ C_{N-n}^{(1+n,1)}}\
\frac{{\rm Vol}\bigl(Fl^{(N)}_{\mathbb R}\bigr)}
     { {\rm Vol}\bigl(Fl^{(n)}_{\mathbb R}\bigr)},
\label{voledgesRe}
\end{equation}
which for $n=1$ gives the volume
$S$ of the boundary $\partial {\cal M}^{\mathbb R}_{N}$,
and allows us to compute the ratio area to volume,
\begin{equation}
\gamma( {\cal M}_{N}^{\mathbb R}) :=
 \frac{\mbox{vol} \bigl( \partial {\cal M}^{\mathbb R}_{N}  \bigr)}
{ {\mbox{vol}}    \bigl( {\cal M}^{\mathbb R}_{N}  \bigr)}  =
\frac{N! \sqrt{N-1}}{\sqrt{N} (N-1)!}\
\frac{C_N^{(1,1)} } {C_{N-1}^{(2,1)}}=
\sqrt{N(N-1)}(N-1)(1+N/2).
\label{muMNRe}
 \end{equation}

The product of the last two factors is equal to the dimensionality
of the set of real density matrices, $D_1=N(N+1)/2-1$. Therefore, just
as in the complex case, the ratio area to volume for
${\cal M}^{\mathbb R}_{N}$ coincides with such a
ratio $\gamma=D_1/L$ for the maximal ball of radius
$L=r_N=[N(N-1)]^{-1/2}$ contained in this set.
In the simplest case of $N=2$ we receive
$V^{(1)}_2=\pi/2$ - the volume of the circle
${\bf B}^2$ of radius $R_2=1/\sqrt{2}$.
The volume of the boundary, $S=\pi\sqrt{2}$,
equals to the circumference of the circle of radius $R_2=1/\sqrt{2}$,
and gives $\gamma=2\sqrt{2}$ in agreement with (\ref{muMNRe}).

\section{Concluding remarks}

We have found the volume $V$ and the surface area $S$ of the
$D=N^2-1$ dimensional set of mixed states ${\cal M}_{N}$ acting in the 
$N$--dimensional Hilbert space,
and its subset ${\cal M}^{\mathbb R}_{N}$ containing real
symmetric matrices. Although the volume of the unitary (orthogonal)
group depends on the definition used, as discussed in the appendix,
the volume of the set of mixed state has a well specified,
unambiguous meaning. For instance, for $N=2$ the volume $V_2$ may be
interpreted as the ratio of the volume of the Bloch ball (of radius
$R_2$ fixed by the Hilbert--Schmidt metric), to the cube
spanned by three orthonormal vectors of the HS space:
the rescaled Pauli matrices,
$\{\sigma_x, \sigma_y, \sigma_z \}/{\sqrt{2}}$.

On one hand, these explicit results may be applied for estimation
of the volume of the set of entangled states
\cite{ZHSL98,Zy99,Sl00,Sl02,Sl03}, or yet
another subset of ${\cal M}_{N}$.
It is also likely to expect that some integrals
obtained in this work will be usefull is such investigations.

On the other hand, outcomes of this paper advance our
understanding of the properties of
the set of mixed quantum states. The ratio
of the hyperarea of the boundary of  $D$--balls to their volume
grows linearly with the dimension $D$. The same ratio
for $D$--simplices behaves as $D^2$, while for the sets
of complex and real density matrices it grows with the dimensionality
$D$ as $D^{3/2}$. Hence these geometrical properties of the convex body of 
mixed states are somewhere in between the properties of
$D$--balls and $D$--simplices.

Furthermore, we have shown that for any $N$
 the sets of complex
(real) density matrices belong to the family
of sets, for which the ratio area to volume is equal to
such a ratio computed for the maximal ball inscribed into this set.

It is necessary to emphasize, that a similar problem of
estimating the volume of the set of mixed states could be also
considered with respect to other probability measures.
In particular, analogous
results presented by us in \cite{SZ03} for the
measure \cite{Ha98,BS01} related
to the Bures distance \cite{Bu69,Uh76}
allow us to investigate similarities and differences
between the geometry of mixed states induced by
different metrics.


It is a pleasure to thank M. Ku{\'s} for helpful discussions
and to P. Slater for valuable correspondence.
Financial support by Komitet Bada{\'n} Naukowych in Warsaw under
the grant 2P03B-072~19 and the Sonder\-forschungs\-be\-reich/Transregio 
12
der Deutschen Forschungs\-gemein\-schaft is gratefully acknowledged.

\appendix

\section{Volumes of the unitary groups and flag manifolds}

Although the volume of the unitary (orthogonal) group and the complex
(real) flag manifold, we use in our calculations, were computed by Hua
many years ago \cite{Hu63}, one may find in more recent literature
related results, which in some cases seem to be contradicting.
However,
different authors used different definitions of the volume
of unitary group \cite{Ma81,Tu85,Fu01,TBS02,BST02,Ca02}, so we
review in this appendix three most common
definitions and compare the results.

\subsection{Unitary group $U(N)$}

We shall recall (\ref{ds2}) the  metric of the unitary group $U(N)$ induced by
 the Hilbert-Schmidt scalar product and used by Hua \cite{Hu63}
\begin{equation}
({\rm d}s)^2  := -{\rm Tr} (U^{-1} {\rm d}U)^2=
\sum_{j=1}^N|(U^{-1}{\rm d}U)_{jj}|^2 +
2 \sum_{j<k=1}^N |(U^{-1}{\rm d}U)_{jk}|^2\ ,
\label{dUs}
\end{equation}
which is left- and right-invariant under unitary transformations. 
The volume element is then given by the product of independent 
differentials times the square root of the determinant of the metric tensor.
 One has still the freedom of an overall scale factor for (\ref{dUs}) 
which appears then correspondingly in the volume element. 
To keep invariance the ratio of the prefactors 
$c_{\rm diag}$ and $c_{\rm off}$  of the diagonal and 
off--diagonal terms has to be fixed. Nevertheless one may introduce 
different scalings of the volume elements which we call $d\nu_A, d\nu_B, d\nu_C$:
\begin{equation}
{\rm d}\nu_A  :=
 \Bigl| \prod_{i=1}^N (U^{-1}{\rm d}U)_{ii} \prod_{j<k}^{1\cdots N}
    \sqrt{2}\ {\rm Re}(U^{-1}{\rm d}U)_{jk}
       {\rm Im}(U^{-1}{\rm d}U)_{jk} \Bigr|;
\quad \quad c_{\rm diag}=1, \quad  c_{\rm off}=2; 
\label{dnuA}
\end{equation}
\begin{equation}
{\rm d}\nu_B :=   2^{-N(N-1)/2}\ {\rm d}\nu_A;
\quad \quad c_{\rm diag}=1, \quad  c_{\rm off}=1;  
\label{dnuB}
\end{equation}
\begin{equation}
{\rm d}\nu_C :=   2^{-N/2}\ {\rm d}\nu_B;
\quad \quad c_{\rm diag}=1/2 \quad  c_{\rm off}=1. 
\label{dnuC}
\end{equation}
The product in (\ref{dnuA}),
consistent with (\ref{dUs}),
 has to be understood in the sense
 of alternating external multiplication of differential forms.
Only the first convention 
(\ref{dnuA}) labeled by the index $_A$ was used in the main part of 
this work. Note that the normalisation (\ref{dnuC})
corresponds to the rescaled line element
$({\rm d}s)^2  = -\frac{1}{2}{\rm Tr} (U^{-1} {\rm d}U)^2$.
In general we may scale
\begin{equation}
{\rm d}\nu_{_X} :=   (c_{\rm diag})^{N/2} (c_{\rm off})^{N(N-1)/2} {\rm d}\nu_B
\label{dnu}
\end{equation}
where the label $_X$ denotes a certain choice  of the prefactors 
$c_{\rm diag}$ and $c_{\rm off}$ for 
diagonal or off--diagonal elements in (\ref{dUs}).
 All these volumes correspond to the
 Haar measure which is unique up to an overall constant scale factor.
Thus we deduce that
\begin{equation}
 {\rm Vol}_A \bigl[ U(N)\bigr] =
 2^{N(N-1)/2}{\rm Vol}_B \bigl[ U(N)\bigr]
{\rm \quad and \quad}
 {\rm Vol}_C \bigl[ U(N)\bigr] =
 2^{-N/2}{\rm Vol}_B \bigl[ U(N)\bigr] .
\label{volUABC}
\end{equation}

In order to determine the volume of the unitary group let us
recall the fiber bundle structure
$U(N-1) \to U(N) \to  {\bf S}^{2N-1}$, see e.g. \cite{Bo91}.
This topological fact implies a relation between
the volume of the unit sphere  ${\bf S}^{2N-1}$ and the volume
of the unitary group defined by the measure $d\nu_B$ (\ref{dnuB}),
for which all components of the vector
 $(U^{-1}{\rm d}U)_{jk}$ have unit prefactors,
\begin{equation}
{\rm Vol}_B[U(N)]= {\rm Vol}_B[U(N-1)] \times {\rm Vol}[{\bf S}^{2N-1}].
\label{volBBS}
\end{equation}
To prove this equality by a direct calculation it is convenient to
parametrize a unitary matrix of size $N$ as
\begin{equation}
U_N=\left[ \begin{array}  [c]{cc}
e^{i\phi}  & 0  \\
0 &  U_{N-1}
\end{array} \right]
\left[ \begin{array}
[c]{cc}
\sqrt{1-|h|^2} &  -h^{\dagger} \\
h & \sqrt{ {\mathbb I} - h\otimes h^{\dagger} }
\end{array} \right] \    ,
  \label{UUUN1}
\end{equation}
 where $\phi \in [0,2\pi)$ is an arbitrary phase
and $h$ is a complex vector with $N-1$ components
such that $|h|\le 1$. This representation shows (we may arrange the two matrices in (\ref{UUUN1}) also in the opposite order) the relation
\begin{equation}
  U(N)/[U(1) \times U(N-1)] = {\mathbb C}{\bf P}^{N-1}\ ,
\label{CPN-1}
\end{equation}
since the second factor represents the complex 
projective space ${\mathbb C}{\bf P}^{N-1}$.
 In fact, if one calculates the metric (\ref{dUs}) we find
\begin{equation}
  ({\rm d}s_N)^2 \cong ({\rm d}s_1)^2 +
  ({\rm d}s_{N-1})^2 + 2({\rm d}s_h)^2
\label{dsN2}
\end{equation}
where $({\rm d}s_N)^2$ means the metric for $U(N)$ 
(the sign $\cong$ shall indicate that we have omitted some 
shifts in $({\rm d}s_1)^2$ and $({\rm d}s_{N-1})^2$ that are not relevant
for the volume) and $({\rm d}s_h)^2$ means the metric of the complex 
projective space ${\mathbb C}{\bf P}^{N-1}$ with radius $1$:
 \begin{equation}
  ({\rm d}s_h)^2 = {\rm d}h^{\dagger}{\rm d}h
   +\frac{(h^{\dagger}{\rm d}h
 +{\rm d}h^{\dagger}h)^2}{4(1-|h|^2)}+
 \frac{ (h^{\dagger}{\rm d}h -{\rm d}h^{\dagger}h)^2}{4}\ .
 \label{dsh2}
 \end{equation}
It is easy to see by diagonalizing this metric 
(eigenvalues $1-|h|^2$, $1/(1-|h|^2)$, and otherwise $1$) that the
 corresponding volume is that of the real ball ${\bf B}^{2N-2}$ 
with radius $1$ and dimension $2N-2$. Thus one obtains
\begin{equation}
{\rm Vol}_X [U(N)]= {\rm Vol}_X [U(N-1)] \times
 {\rm Vol}_X [U(1)] \times  c_{\rm off}^{N-1}
{\rm Vol}[{\bf B}^{2N-2}]  \  ,
\label{volBBS2}
\end{equation}
which for the measure (\ref{dnuB}) with
$c_{\rm diag}=c_{\rm off}=1$ reduces to
(\ref{volBBS}). Applying this relation  $N-1$ times we obtain
\begin{equation}
 {\rm Vol}_B \bigl[U(N)]=
 {\rm Vol}[{\bf S}^{2N-1}]  \times \cdots \times
 {\rm Vol}[{\bf S}^{3}] \times {\rm Vol}[{\bf S}^{1}] .
\label{volUNB}
\end{equation}
Taking into account that Vol$[{\bf S}^{2N-1}]=2\pi^N/(N-1)!$
and making use of the relation (\ref{volUABC})
we may write an explicit result for the volumes calculated
with respect to different definitions (\ref{dnuA} -- \ref{dnuC})
\begin{equation}
 {\rm Vol}_X  [U(N)]=
 a^U_X  \   \frac {2^N \pi^{N(N+1)/2}}{0!1!\cdots (N-1)!},
\label{volUBB}
\end{equation}
where the proportionality constants read $a^U_A=2^{N(N-1)/2}$,
$a^U_B=1$ and $a^U_C=2^{-N/2}$. The result for ${\rm Vol}_A[U(N)]$
was rigorously derived in \cite{Hu63}, 
${\rm Vol}_B[U(N)]$ was given in \cite{Fu01},
while ${\rm Vol}_A[U(N)]$ and ${\rm Vol}_C[U(N)]$
were compared in \cite{Ca02}.
In particular, ${\rm Vol}_A[U(1)]={\rm Vol}_B[U(1)]=2\pi$, while
${\rm Vol}_C[U(1)]={\sqrt{2}}\pi$ and ${\rm Vol}_A[U(2)]=8\pi^3$,
${\rm Vol}_B[U(2)]=4\pi^3$, ${\rm Vol}_C[U(2)]=2\pi^3$.

In general,  the volume of
 a coset space may be expressed as a ratio of the volumes.
Consider for instance the manifold of all pure states of dimensionality
$N$. It forms the complex projective space
 ${\mathbb C}{\bf P}^{N-1}=U(N)/[U(N-1) \times U(1)]$.
Therefore
\begin{equation}
{\rm Vol}_X [ {\mathbb C}{\bf P}^{N-1} ]
 = \frac {{\rm Vol}_X [U(N)] } { {\rm Vol}_X [U(1)]\ {\rm Vol}_X [U(N-1)]} \ ,
\label {CPNIND}
\end{equation}
which gives the general result
\begin{equation}
 {\rm Vol}_X [{\mathbb C}{\bf P}^{k}]
  = a^{\rm CP}_X \   \frac {\pi^{k}}{k!}
  = a^{\rm CP}_X  \  {\rm Vol} [ {\bf B}^{2k} ]
\label{volcpn}
\end{equation}
The scale factors read $a^{\rm CP}_A=2^k$ and $a^{\rm CP}_B=a^{\rm CP}_C=1$.
For instance ${\rm Vol}_A [{\mathbb C}{\bf P}^{1}]=2\pi$
which corresponds to the circle of radius $ \sqrt{2}$  ,
while  ${\rm Vol}_B [{\mathbb C}{\bf P}^{1}]=
{\rm Vol}_C [{\mathbb C}{\bf P}^{1}]=\pi$,
equal to the area of the circle of radius $1 $. The latter convention is
natural if one uses the Fubini--Study metric in the space of pure states,
$D_{FS}(|\varphi\rangle,|\psi\rangle) = {\rm arccos}(\sqrt{\kappa})$,
where the transition probability is given by $\kappa = | \langle \varphi | \psi \rangle|^2$.
Then the largest possible distance $D_{FS}=\pi/2$, obtained for any orthogonal states,
sets the geodesic length of the complex projective space to $\pi$ which corresponds to the geodesic distance of two opposite points on the unit circle, being identified. It is worth to add
that ${\rm Vol}_C [{\mathbb C}{\bf P}^{k}]={\rm Vol}[ {\bf S}^{2k+1}] / {\rm Vol} [{\bf S}^{1}]$
and this relation was used in \cite{BST02} to {\sl define} the volume
Vol$_C$ of complex projective spaces.
We see therefore that different conventions adopted in (\ref{dnuA} -- \ref{dnuC})
lead to different sizes (geodesic lengths) of the manifolds analyzed.

Unitary orbits  of a generic mixed state with a
non--degenerate spectrum
have structure of a $(N^2-N)$--dimensional complex flag manifold
$Fl^{(N)}_{\mathbb C}=U(N)/[U(1)]^N$. Hence its volume reads
\begin{equation}
 {\rm Vol}_X [ Fl_{\mathbb C}^{(N)}]  =
\ \frac {  {\rm Vol}_X [ U(N)] }
         { \bigl( {\rm Vol}_X [U(1)] \bigr)^N} =
a_X^{\rm Fl} \  \frac { \pi^{N(N-1)/2}}{1!2!\cdots (N-1)! }
\label{volflagC}
\end{equation}
with convention dependent scale constants
$a^{\rm Fl}_A=2^{N(N-1)/2}$ \cite{Hu63} and $a^{\rm Fl}_B=a^{\rm Fl}_C=1$ \cite{BST02}.
It is easy to check that the relation
\begin{equation}
 {\rm Vol}_X [ Fl_{\mathbb C}^{(N)}]  =
 {\rm Vol}_X [{\mathbb C}{\bf P}^{1}]  \times
 {\rm Vol}_X [{\mathbb C}{\bf P}^{2}] \times \cdots \times
  {\rm Vol}_X [{\mathbb C}{\bf P}^{N-1}]
 \label{volflagC2}
\end{equation}
holds  for any definition (\ref{dnuA} -- \ref{dnuC}),
since the scale constants do cancel.

For completeness we discuss also the group $SU(N)$, the volume of which is {\sl not}
equal to  ${\rm Vol}[U(N)] /  {\rm Vol} [U(1)]$ \cite{Ma81,TBS02,BST02}.
 To show this let us parametrize
a matrix $Y_N \in SU(N)$
v\begin{equation}
Y_N=\left[ \begin{array}  [c]{cc}
e^{i\phi}  & 0  \\
0 & e^{-i[\phi/(N-1)]} Y_{N-1}
\end{array} \right]
\left[ \begin{array}
[c]{cc}
\sqrt{1-|h|^2} &  -h^{\dagger} \\
h & \sqrt{ {\mathbb I} - h\otimes h^{\dagger} }
\end{array} \right]   =VW \ ,
  \label{SUN1}
\end{equation}
 where $\phi \in [0,2\pi)$ is an arbitrary phase
and $h$ is a complex vector with $N-1$ components
such that $|h|\le 1$.  Condition det$Y_{N}=1$ implies
Tr$Y^{-1}_N{\rm d}Y_N=0.$
For instance, the metric  (\ref{dUs}) gives, if the volume is concerned
\begin{equation}
({\rm d}s)^2 \cong  -{\rm Tr} (V^{-1} {\rm d}V)^2 -
  {\rm Tr} (W^{-1} {\rm d}W)^2 .
\label{dSUNb}
\end{equation}
Since the first factor $V$ is block diagonal
the first term is equal to
$({\rm d}\phi)^2 N/(N-1) - {\rm Tr} (Y_{N-1}^{-1} {\rm d} Y_{N-1})^2$, while
the second one gives the metric on ${\mathbb C}{\bf P}^{N-1}$.
Integrating an analogous expression in the general case of an arbitrary metric
and using (\ref {CPNIND}) we obtain the following result
\begin{equation}
 {\rm Vol}_X \bigl[SU(N)]=
 \frac { {\rm Vol}_X [U(N)]} { {\rm Vol}_X [U(N-1)]}
 \ \sqrt{\frac {N}{N-1}}
 \ {\rm Vol}_X \bigl[SU(N-1)] ,
\label{volSUN1}
\end{equation}
which iterated $N-1$ times gives the correct relation
\begin{equation}
 {\rm Vol}_X \bigl[SU(N)] =
 \sqrt{N} \ \frac { {\rm Vol}_X [U(N)]} { {\rm Vol}_X [U(1)]}
\label{volSUN2}
\end{equation}
with the stretching factor $\sqrt{N}$.
For instance, working with the measure (\ref{dnuC})
and making use of (\ref{volUBB}) we obtain
${\rm Vol}_C \bigl[SU(N)]=\sqrt{N}2^{(N-1)/2}\pi^{(N+2)(N-1)/2}
/[1!\cdots (N-1)!]$, so in particular,  
$ {\rm Vol}_C \bigl[SU(2)]=2\pi^2$,
$ {\rm Vol}_C \bigl[SU(3)]=\sqrt{3} \pi^5$ and
$ {\rm Vol}_C \bigl[SU(4)]=\sqrt{2}\pi^9/3$ in consistence with
results obtained  in \cite{Ma81,By98,BST02,Ca02}.

\subsection{Orthogonal group $O(N)$}

The analysis of the orthogonal group is simpler,
since  $(O^{-1} {\rm d}O)^T=- (O^{-1} {\rm d}O)^T$, so
the diagonal elements of ${\rm Tr} (O^{-1} {\rm d}O)^2$  vanish.
Thus we shall consider only two metrics
(analogous to the measures (\ref{dnuA} -- \ref{dnuC})) with different scalings,
\begin{equation}
({\rm d}s_A)^2  := -{\rm Tr} (O^{-1} {\rm d}O)^2=
2 \sum_{j<k=1}^N |(O^{-1}{\rm d}O)_{jk}|^2,
\label{dOsa}
\end{equation}
used in section VII of this work, and
\begin{equation}
({\rm d}s_B)^2= ({\rm d}s_C)^2
:= -\frac{1}{2}{\rm Tr} (O^{-1} {\rm d}O)^2=
 \sum_{j<k=1}^N |(O^{-1}{\rm d}O)_{jk}|^2 ,
\label{dOsb}
\end{equation}
which both lead to the Haar measure on the orthogonal group.

To obtain the volume of $O(N)$
we proceed as in the unitary case and
parametrize an orthogonal matrix of size $N$ as
\begin{equation}
O_N=\left[ \begin{array}  [c]{cc}
O_1  & 0  \\
0 &  O_{N-1}
\end{array} \right]
\left[ \begin{array}
[c]{cc}
\sqrt{1-|h|^2} &  -h^{T} \\
h & \sqrt{ {\mathbb I} - h \otimes  h^{T} }
\end{array} \right] \    ,
  \label{OOON1}
\end{equation}
 where $O_1\in O(1)=\pm $, while  $h$ is here
a real vector with $N-1$ components
such that $|h|\le 1$.
Representing the metric $({\rm d}s_B)^2$ by these two matrices
we see that the term containing only the vector $h$
gives the metric of a real projective space.
Integrating the resulting volume element (with scale factor $1$) we
obtain the volume of ${\mathbb R}{\bf P}^{N-1}$, equal to
$\frac{1}{2} {\rm Vol} [{\bf S}^{N-1}]$. Taking into account a factor of two
resulting from $O(1)$ we arrive at
Vol$_B[O(N)]={\rm Vol}_B [O(N-1)] {\rm Vol} [{\bf S}^{N-1}]$,
which applied recursively leads to
\begin{equation}
  {\rm Vol}_B[O(N)]=
  {\rm Vol} [{\bf S}^{N-1}] \times \cdots \times
  {\rm Vol} [{\bf S}^{1}] \times
  {\rm Vol} [{\bf S}^{0}] = \prod_{k=1}^N \frac{2 \pi^{k/2}}{\Gamma(k/2)} ,
\label{volON1}
\end{equation}
where ${\rm Vol} [{\bf S}^{0}] = {\rm Vol} [O(1)] = 2$.
To get an equivalent result for the metric (\ref{dOsa})
we have to take into account the factor $\sqrt{2}$
which occures for each of $N(N-1)/2$ off-diagonal elements.
Doing so we obtain
\begin{equation}
  {\rm Vol}_A[O(N)]=
 2^{N(N-1)/4} {\rm Vol}_B[O(N)]=
 2^{N(N+3)/4}  \prod_{k=1}^N \frac{ \pi^{k/2}}{\Gamma(k/2)} ,
\label{volON2}
\end{equation}
in consistency with Hua \cite{Hu63}.
In particular, ${\rm Vol}_A[O(1)]={\rm Vol}_B[O(1)]=2$, while
${\rm Vol}_A[O(2)]=4\sqrt{2} \pi$, ${\rm Vol}_B[O(2)]=4\pi$,
and ${\rm Vol}_A[O(3)]=32\sqrt{2}\pi^2$, ${\rm Vol}_B[O(3)]=16\pi^2$.

In full analogy to the unitary case we obtain the volume
of the real projective manifold
\begin{equation}
{\rm Vol}_X [ {\mathbb R}{\bf P}^{N-1} ]
 = \frac {{\rm Vol}_X [O(N)] } { {\rm Vol}_X [O(1)]\ {\rm Vol}_X [O(N-1)]} \ .
\label {RPNIND}
\end{equation}
For the metric (\ref{dOsb})
this expression reduces to
${\rm Vol}_B [ {\mathbb R}{\bf P}^{k} ]=\frac{1}{2}{\rm Vol} [{\bf S}^{k}]$.
Hence, this metric may be called  'unit sphere' metric,
while the convention (\ref{dOsa})
may be called 'unit trace' metric.

In the similar way we find the  volume of the real flag manifolds, used in analysis
of real density matrices,
\begin{equation}
 {\rm Vol}_X [ Fl_{\mathbb R}^{(N)}]  =
\ \frac {  {\rm Vol}_X [ O(N)] }
         { \bigl( {\rm Vol}_X [O(1)] \bigr)^N} =
\ \frac {1}{2^N}  {\rm Vol}_X [ O(N)]  .
\label{volflagR}
\end{equation}
Exactly as in the complex case we observe that the relation
\begin{equation}
 {\rm Vol}_X [ Fl_{\mathbb R}^{(N)}]  =
 {\rm Vol}_X [{\mathbb R}{\bf P}^{1}]  \times
 {\rm Vol}_X [{\mathbb R}{\bf P}^{2}] \times \cdots \times
  {\rm Vol}_X [{\mathbb R}{\bf P}^{N-1}]
 \label{volflagR2}
\end{equation}
is satisfied for any definition of the metric.

Computation of the volume of the special orthogonal group $SO(N)$
is much easier than in the complex case, since there are no diagonal elements
in the metric and hence, no stretching factors. For any normalization
 one gets
\begin{equation}
{\rm Vol}_X [SO(N)]
 = \frac {{\rm Vol}_X [O(N)] } { {\rm Vol}_X [O(1)] } =
\frac{1}{2}  \ {\rm Vol}_X [O(N)]
\label {SON}
\end{equation}
In particular,  we get
$ {\rm Vol}_B [SO(2)]=2\pi$ and
$ {\rm Vol}_B [SO(3)]= {\rm Vol}_C \bigl[SO(3)]=8\pi^2$.
The latter results seem to be inconsistent with
$ {\rm Vol}_C [SU(2)]=2\pi^2$, since there exists a one to two relation between
both groups. This paradox is resolved by analyzing the scale effects \cite{BST02}:
the volume of $SU(2)$ is two times larger than
the volume of the real projective manifold conjugated to $SO(3)$
of the appropriate geodesic length,
$ {\rm Vol}_C [{\mathbb R}{\bf P}^3]=\pi^2$.

\end{document}